\documentclass[manuscript, screen, authorversion]{acmart}

\AtBeginDocument{%
  }

\copyrightyear{2025}
\acmYear{2025}
\setcopyright{rightsretained}
\acmConference[AutomotiveUI Adjunct '25]{17th International Conference on Automotive User Interfaces and Interactive Vehicular Applications}{September 21--25, 2025}{Brisbane, QLD, Australia}
\acmBooktitle{17th International Conference on Automotive User Interfaces and Interactive Vehicular Applications (AutomotiveUI Adjunct '25), September 21--25, 2025, Brisbane, QLD, Australia}
\acmDOI{10.1145/3744335.3758497} 
\acmISBN{979-8-4007-2014-7/2025/09}

\acmSubmissionID{44}

\usepackage[T1]{fontenc}
\usepackage{xcolor}
\usepackage{xspace}
\usepackage[capitalise]{cleveref}
\usepackage{subcaption,amsfonts,dcolumn} 
\usepackage{gensymb} 

\newcommand{\ie}{\emph{i.e.}\xspace}
\newcommand{\eg}{\emph{e.g.}\xspace}

\newcommand{\aka}{\emph{a.k.a.}\xspace}

\newcommand{\LevelOne}{Level~1\xspace}
\newcommand{\LevelTwo}{Level~2\xspace}
\newcommand{\LevelThree}{Level~3\xspace}

\begin{document}

\title[Gaze-Based Indicators of Driver Cognitive Distraction]{Gaze-Based Indicators of Driver Cognitive Distraction: Effects of Different Traffic Conditions and Adaptive Cruise Control Use}

\author{Ana{\"i}s Halin}
\email{anais.halin@uliege.be}
\orcid{0000-0003-3743-2969}
\affiliation{%
  \institution{University of Liège}
  \department{Department of Electrical Engineering and Computer Science}
  \city{Liège}
  \country{Belgium}
}

\author{Adrien Deli{\`e}ge}
\email{adrien.deliege@uliege.be}
\orcid{0000-0003-3981-6982}
\affiliation{%
  \institution{University of Liège}
  \department{Department of Electrical Engineering and Computer Science}
  \city{Liège}
  \country{Belgium}
}

\author{Christel Devue}
\email{cdevue@uliege.be}
\orcid{0000-0001-7349-226X}
\affiliation{%
  \institution{University of Liège}
  \department{Department of Psychology}
  \city{Liège}
  \country{Belgium}
}

\author{Marc Van Droogenbroeck}
\email{m.vandroogenbroeck@uliege.be}
\orcid{0000-0001-6260-6487}
\affiliation{%
  \institution{University of Liège}
  \department{Department of Electrical Engineering and Computer Science}
  \city{Liège}
  \country{Belgium}
}


\begin{abstract}
In this simulator study, we investigate how gaze parameters reflect driver cognitive distraction under varying traffic conditions and adaptive cruise control (ACC) use.
Participants completed six driving scenarios that combined two levels of cognitive distraction (with/without mental calculations) and three levels of driving environment complexity. Throughout the experiment, participants were free to activate or deactivate an ACC.
We analyzed two gaze-based indicators of driver cognitive distraction: the percent road center, and the gaze dispersions (horizontal and vertical). 
Our results show that vertical gaze dispersion increases with traffic complexity, while ACC use leads to gaze concentration toward the road center. Cognitive distraction reduces road center gaze and increases vertical dispersion. Complementary analyses revealed that these observations actually arise mainly between mental calculations, while periods of mental calculations are characterized by a temporary increase in gaze concentration.
\end{abstract}

\ccsdesc[500]{Human-centered computing~Empirical studies in HCI}
\ccsdesc[500]{Human-centered computing~Laboratory experiments}

\keywords{Driver State, Cognitive Distraction, Workload, Indicators, Gaze Parameters, Percent Road Center, Gaze Dispersion, Simulator Study, Driving Automation, Adaptive Cruise Control, Driving Environment Complexity, Traffic Conditions}

\maketitle

\section{Introduction}
Monitoring the state of drivers that are actively controlling a vehicle, supervising driving automation features, or expected to take over at a given moment, is a critical safety concern to ensure that they are ready to perform the required actions.
Cognitive distraction, a state in which drivers' mental resources are diverted from the driving task (\eg, talking with a passenger or on the phone), can significantly impair performance and jeopardize safety. Monitoring behavioral signals, such as gaze parameters, can be used to detect this state. 
However, such signals may also be influenced by external factors such as traffic conditions and the use of driving automation features like adaptive cruise control (ACC). 

We investigate how gaze parameters reflect driver cognitive distraction under varying traffic conditions and ACC use.
We analyzed data collected during a driving simulator study~\cite{Halin2025Effects}, in which participants could freely turn the ACC on or off while driving across six scenarios. These scenarios varied along two dimensions: driver state, with two levels of cognitive distraction (\ie, presence/absence of a mental calculation task), and driving environment complexity, with three levels (\ie, increasing traffic density and addition of road constructions restricting the number of traffic lanes).

We aim at answering the following research question: \emph{What are the effects of cognitive distraction on gaze parameters under varying traffic conditions and ACC use?}

\section{Related Work}
Driver state can be categorized into five (sub-)states: drowsiness, mental workload, distraction (manual, visual, auditory, and cognitive), emotions, and under influence~\cite{Halin2021Survey}. 
Driver monitoring systems aim to characterize these states using specific indicators and appropriate sensors (to access the values of these indicators).
\citet{Halin2021Survey} provide a comprehensive review of the indicators and sensors that support the characterization of all these states.

Cognitive distraction, defined by the National Highway Traffic Safety Administration (NHTSA)~\cite{NHTSA2010Overview} as the mental workload associated with a task that involves thinking about something else apart from the driving task, is commonly characterized using behavioral and physiological indicators, such as gaze parameters (\eg, fixation duration, gaze distribution)~\cite{Strayer2015Assessing, Marquart2015Review, Le2020Evaluating}, pupil diameter~\cite{Yokoyama2018Prediction}, electrodermal activity (EDA)~\cite{Yusoff2017Selection}, and heart activity~\cite{Paxion2014Mental, Reimer2009AnOnroad}. Gaze parameters are particularly attractive, as they are non-intrusive and easily collected in real-world driving conditions.

In the driving context, visual and mental tasks are inherently intertwined. Drivers must continuously perceive their environment and interpret visual information to make appropriate decisions, such as, \eg, scanning highway lanes while simultaneously determining when to brake or change lanes.
Given this strong link between visual perception and cognitive processing, gaze parameters have often been used to assess drivers' mental workload~\cite{Marquart2015Review}.

Different approaches exist to code glance data, namely the direction-based (where are drivers looking?), the target-based (what objects are drivers looking at?), and the purpose-based (why are drivers looking where they are looking?) coding schemes of glance data~\cite{Ahlstrom2021Eye,Kircher2024AComparison}. 
Direction-based coding is commonly used to compute indicators like ``eyes off road'' or ``percent road center'', which is the percentage of gaze data points that fall within a predefined road center area~\cite{Victor2005Sensitivity}. 
Target-based coding involves a manual coding of glance targets, which are categorized by the target type (\eg, bicyclist, traffic sign, or mobile phone).
The purpose-based approach specifically defines which areas a driver must acquire information from to be considered attentive~\cite{Ahlstrom2021Eye}. 
 
Target-based approaches infer driver distraction from glances toward targets deemed irrelevant for driving. 
While effective for detecting visual distraction, they are less suited for cognitive distraction, where drivers may maintain their gaze on the road while being mentally distracted. 
By definition, purpose-based approaches might better capture cognitive distraction but still face similar limitations, as drivers may appear visually attentive while mentally disengaged (\ie, ``look but not see'' situations, where drivers' eyes are on the road, but the mind is elsewhere). Moreover, their practical implementation is complex as it requires extensive contextual information about, \eg, glance history, infrastructure layout, or traffic regulations.
Therefore, we focus on direction-based gaze indicators of cognitive distraction, which are more accessible and widely used. Specifically, we examine percent road center, horizontal and vertical gaze dispersion. 

In that vein, cognitive distraction has been linked with increased gaze concentration toward the road center, manifested as higher percent road center~\cite{Victor2005Sensitivity}, lower horizontal and vertical gaze dispersion, and decreased glance frequency to mirrors and speedometer~\cite{Victor2005Sensitivity,Harbluk2007AnOnroad}. 
Similar patterns have also been noted under high driving task complexity~\cite{Victor2005Sensitivity}.

Let us note that percent road center areas can vary considerably across studies: they are sometimes circular~\cite{Victor2005Sensitivity,Ahlstrom2009Considerations,Wang2014TheSensitivity}, sometimes rectangular~\cite{Harbluk2007AnOnroad,Victor2005Sensitivity}, with horizontal angle of view ranging from $8\degree$ to $20\degree$.
Furthermore, this area can be a fixed area on the windshield or can be centered around the driver’s most frequent gaze angle, and thus vary for each driver~\cite{Wang2014TheSensitivity}.
While percent road center was initially computed based on gaze fixations, using raw gaze points (\ie, all gaze data not clustered into fixations and saccades) was shown to give highly correlated values~\cite{Ahlstrom2009Considerations,Wang2014TheSensitivity}.

Gaze dispersion, expressed as the standard deviation of gaze points (\aka, gaze variability), can be computed using the driver’s gaze angle or the projection of the gaze trail on a plane in front of the driver~\cite{Victor2005Sensitivity}. Furthermore, it can also be assessed separately for the horizontal and vertical gaze positions or angles~\cite{Wang2014TheSensitivity}, or for the combined vertical and horizontal components of gaze into a single gaze position or angle~\cite{Victor2005Sensitivity}.

In this study, we compute (1) the percent road center using raw gaze angles and a rectangular road center area with an horizontal angle of view of $20\degree$; and (2) horizontal and vertical dispersion of gaze angles.

\section{User Study}
This study is based on data collected during a driving simulator experiment designed to investigate how drivers’ cognitive state and driving environment complexity influence their reliance on driving automation features. Full details regarding the materials, experimental tasks, and procedure can be found in the paper of~\citet{Halin2025Effects}.

This within-subjects study investigates how the presence of a cognitively distracting task affects gaze parameters, considering variations in both traffic complexity and ACC use.

\subsection{Participants}
Thirty-one individuals were initially recruited via posters, social media, and word of mouth. Two withdrew due to simulator sickness, resulting in a final sample of $N=29$ participants ($22$ male, $7$ female). All held valid driver’s licenses.
The study received ethics approval from the Faculty of Psychology, Logopedics and Educational Sciences of the University of Li{\`e}ge, Belgium (ref. 2223‑081). 
All participants gave written informed consent.

\subsection{Apparatus and Materials}
\subsubsection{Driving Simulator}
The study was conducted using a driving simulator developed by AISIN Europe with three $50$-inch curved screens, an adjustable seat, and a Fanatec steering wheel, shifter, and pedals. CARLA~\cite{Dosovitskiy2017CARLA} (an open-source Unreal Engine simulator) allowed flexible sensor/control configuration and scenario scripting.
The vehicle featured intelligent ACC (adapting speed to both leading vehicle and curves), adjustable in $5$~km/h increments via steering‑wheel buttons. A secondary, smaller screen mimicked a navigation display.
A custom software from AISIN Europe handled scenario design, simulation execution, and data logging.

\subsubsection{Driver Monitoring System}
\begin{figure*}
     \centering
     \includegraphics[width=\linewidth]{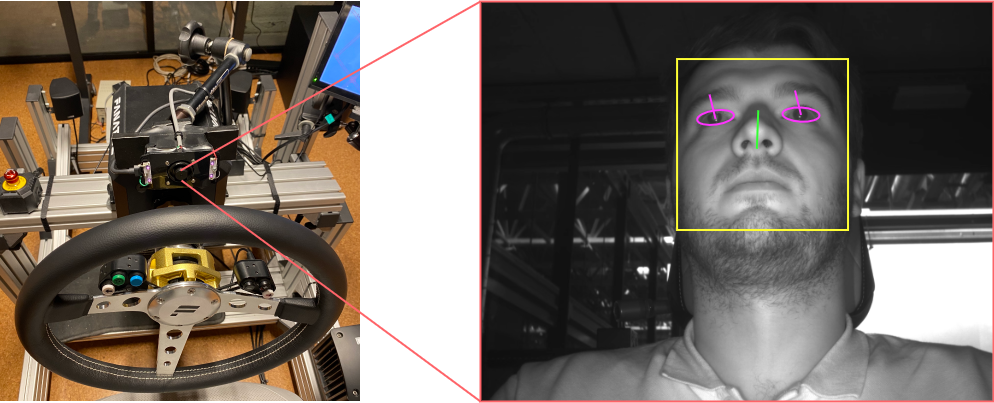}
     \Description{The figure shows on the left the infrared camera positioned behind the wheel and on the right an image acquired by the camera and processed by the algorithms developed by AISIN Europe.}
     \caption{Illustration of the setup for the driver monitoring system. The person depicted provided permission.}
     \label{fig:DMS_setup}
 \end{figure*}

The driver monitoring system was composed of a high-resolution infrared camera positioned behind the wheel (see \cref{fig:DMS_setup}), recording images at $60$ Hz. Images were processed in real time by algorithms developed by AISIN Europe. Face orientation (pitch, yaw, roll, in degrees), eye-opening (in millimeter), and gaze direction (horizontal and vertical gaze angles) were logged. 

\subsection{Procedure and Tasks}
Participants were first equipped with EDA electrodes (for a separate study), then completed a pre-test questionnaire collecting demographic and study-specific information.

After a brief training session, they performed $6$ repetitions of the same driving route, combining $3$ levels of driving environment complexity and $2$ levels of cognitive distraction (with vs. without a secondary task), in a within-subjects design. Each session lasted approximately $8$ minutes. 
The driving environment complexity varied in traffic conditions (low for \LevelOne, and medium for \LevelTwo and \LevelThree) and road construction zones ($0$ for \LevelOne and \LevelTwo, and $3$ for \LevelThree).
The route included urban segments with traffic lights and highway segments, under clear daylight conditions. In the cognitive distraction conditions, participants had to respond orally to two-digit additions (\eg, $83+42$) dictated by a voice agent.
They were instructed to follow the navigation system, comply with traffic rules, including speed limitations, and were free to activate or deactivate the ACC at any time during each drive. The order of scenarios was counterbalanced across participants, with each pair of scenarios at the same complexity level presented consecutively, once with and once without the cognitively distracting task.
Driving environment complexity varied across the three levels in terms of traffic density.

Each scenario began with a brief calibration phase to synchronize all data streams. After each one, participants completed a feedback questionnaire. Analysis of both pre-test and feedback questionnaire data is presented in~\cite{Halin2025Effects}.

\subsection{Measurements}  
We assessed whether two gaze-related indicators correlate with cognitive distraction and how they are influenced by different traffic conditions and ACC use. We analyzed (1)~the percent road center~\cite{Victor2005Sensitivity}, which is the percentage of gaze data points that fall within the road center area, and (2)~the horizontal and vertical gaze dispersions, which are computed as the standard deviations of horizontal and vertical gaze angles, respectively. 

For each participant and each session, the road center point was determined as the mode, or most frequent gaze angle, and the road center area was defined as a $20\degree$ (horizontal) $\times$ $15\degree$ (vertical) rectangular area centered around the road center point, as done in~\cite{Victor2005Sensitivity}. In line with prior work showing high correlation between fixation-based and raw-data-based indicators~\cite{Ahlstrom2009Considerations, Wang2014TheSensitivity}, we used all gaze data points to compute the indicators, without distinguishing between fixations and saccades. However, we analyzed these indicators only during the drive on the highway segment, which lasted approximately $5$ minutes (thus removing the driving period in the city center, where vehicles were often stopped at traffic lights).

\section{Results}
Statistical analyses were conducted using JASP (Version 0.19.3)~\cite{JASP2025}. 
Descriptive and inferential statistics are reported for each indicator. Results were considered statistically significant at $p<.05$.

\subsection{Main Analyses of Gaze-Related Indicators of Cognitive Distraction}
The $29$ participants completed $6$ scenarios ($3$ levels of driving complexity $\times$ $2$ levels of cognitive distraction) in which they could freely activate or deactivate the ACC, allowing for up to $348$ observations ($29 \times 6 \times 2$). However, in $7$ of them, ACC was never used or the gaze direction was not detected at all, resulting in missing data and an unbalanced dataset for analyses involving ACC use, thus preventing us from conducting a standard analysis of variance (ANOVA).
Therefore, we defaulted to a linear mixed model using Restricted Maximum Likelihood (REML) and Satterthwaite's approximation, based on the remaining $341$ observations. The model included environment complexity, cognitive distraction, and ACC use as fixed effects, with a random intercept for participants to account for individual differences.

\subsubsection{Percent Road Center}
There were significant main effects of cognitive distraction ($F(1,301.02)=14.196, p<.001$) and ACC use ($F(1,301.04)=27.587, p<.001$) on percent road center. However, there was no significant effect of driving environment complexity ($F(2,301.03)=1.197, p=.303$), and no significant cross-factor interaction. 
As shown in~\cref{tab:PRC-main}~(a), percent road center was lower when participants were cognitively distracted ($73.3 \pm 12.8$ globally), compared to when they were not ($76.1 \pm 12.8$ globally). Percent road center was also lower when ACC was off ($71.9 \pm 13.8$ globally), compared to when ACC was on ($75.8 \pm 13.4$ globally). Participants were thus looking more often at the road center area during sessions without distraction and when the ACC was activated. 

\begin{table*}[ht]
    \centering
    \caption{\textbf{Average gaze-based indicators of cognitive distraction ($\pm$ standard deviation) for the $N=29$ participants.} Main analyses (tables a, b, c) report values for $12$ conditions: ACC engaged vs. disengaged across $3$ levels of driving environment complexity (DEC, from \LevelOne with low complexity to \LevelThree with high complexity) and $2$ levels of cognitive distraction (no distraction vs. distraction). Complementary analyses (tables a', b', c') focus on the sessions with cognitive distraction at each DEC level, comparing segments where participants performed mental calculations (calculations) with interleaving segments (no calculations).}
    \label{tab:gaze_indicators}
    
    \begin{subtable}{.55\textwidth}
    \centering
    \caption*{(a) \textbf{Percent road center:} Main analysis}
    \label{tab:PRC-main}
    \begin{tabular}{l|c|c|c|c|}
        \cline{2-5}
                                                    & \multicolumn{2}{c|}{\textbf{No distraction}}  & \multicolumn{2}{c|}{\textbf{Distraction}}\\
        \hline
        \multicolumn{1}{|l|}{\textbf{DEC} }         & \textbf{ACC Off}      & \textbf{ACC On}       & \textbf{ACC Off}      & \textbf{ACC On} \\    
        \hline
        \multicolumn{1}{|l|}{\textbf{\LevelOne} (low)}    & $73.3 \pm 13.3$      & $78.9 \pm 13.0$       & $69.1 \pm 14.3$       & $74.9 \pm 14.2$ \\
        \multicolumn{1}{|l|}{\textbf{\LevelTwo} (medium)}    & $72.7 \pm 15.8$       & $78.5 \pm 11.3$       & $72.0 \pm 12.4$       & $74.8 \pm 13.0$ \\
        \multicolumn{1}{|l|}{\textbf{\LevelThree} (high)}  & $72.7 \pm 13.1$       & $75.8 \pm 13.7$       & $71.5 \pm 14.1$       & $72.0 \pm 15.1$ \\
        \hline
    \end{tabular}
    \end{subtable} 
    \hspace{4em}
    \begin{subtable}{.35\textwidth}
    \centering
    \caption*{(a') Complementary analysis}
    \label{tab:PRC-complementary}
    \begin{tabular}{|c|c|}
        \hline
        \multicolumn{2}{|c|}{\textbf{Distraction}}\\
        \hline
        \textbf{Calculations}      & \textbf{No Calculations} \\    
        \hline
        $74.7 \pm 14.9$       & $73.2 \pm 12.7$ \\
        $75.7 \pm 13.5$       & $73.0 \pm 12.6$ \\
        $72.6 \pm 14.4$       & $71.9 \pm 12.9$ \\
        \hline
    \end{tabular}
    \end{subtable} 

    \bigskip

    \begin{subtable}{.55\textwidth}
    \centering
    \caption*{(b) \textbf{Horizontal gaze dispersion:} Main analysis}
    \label{tab:horizontal_gaze_dispersion-main}
    \begin{tabular}{l|c|c|c|c|}
        \cline{2-5}
                                                    & \multicolumn{2}{c|}{\textbf{No distraction}}      & \multicolumn{2}{c|}{\textbf{Distraction}}\\
        \hline
        \multicolumn{1}{|l|}{\textbf{DEC} }         & \textbf{ACC Off}      & \textbf{ACC On}           & \textbf{ACC Off}             & \textbf{ACC On} \\    
        \hline
        \multicolumn{1}{|l|}{\textbf{\LevelOne} (low) }   & $9.0 \pm 2.9$   & $7.1 \pm 2.6$   & $8.5 \pm 3.3$   & $7.2 \pm 2.3$ \\
        \multicolumn{1}{|l|}{\textbf{\LevelTwo}  (medium)}    & $8.5 \pm 3.1$   & $7.1 \pm 2.4$   & $8.7 \pm 2.9$   & $7.6 \pm 2.3$ \\
        \multicolumn{1}{|l|}{\textbf{\LevelThree} (high)} & $9.0 \pm 3.2$   & $7.4 \pm 2.4$   & $8.9 \pm 3.3$   & $7.1 \pm 2.2$ \\
        \hline
    \end{tabular}
    \end{subtable}
    \hspace{4em}
    \begin{subtable}{.35\textwidth}
    \centering
    \caption*{(b') Complementary analysis}
    \label{tab:horizontal_gaze_dispersion-complementary}
    \begin{tabular}{|c|c|}
        \hline
        \multicolumn{2}{|c|}{\textbf{Distraction}}\\
        \hline
        \textbf{Calculations}      & \textbf{No Calculations} \\    
        \hline
        $6.6 \pm 2.1$       & $8.1 \pm 2.5$ \\
        $7.2 \pm 2.7$       & $8.4 \pm 2.4$ \\
        $7.3 \pm 2.5$       & $8.3 \pm 2.5$ \\
        \hline
    \end{tabular}
    \end{subtable} 

    \bigskip

    \begin{subtable}{.55\textwidth}
    \centering
    \caption*{(c) \textbf{Vertical gaze dispersion:} Main analysis}
    \label{tab:vertical_gaze_dispersion-main}
    \begin{tabular}{l|c|c|c|c|}
        \cline{2-5}
                                                & \multicolumn{2}{c|}{\textbf{No distraction}}                & \multicolumn{2}{c|}{\textbf{Distraction}}\\
        \hline
        \multicolumn{1}{|l|}{\textbf{DEC} }     & \textbf{ACC Off}            & \textbf{ACC On}               & \textbf{ACC Off}             & \textbf{ACC On} \\    
        \hline
        \multicolumn{1}{|l|}{\textbf{\LevelOne} (low)}    & $6.3 \pm 2.0$   & $6.0 \pm 2.2$    & $7.2 \pm 2.8$   & $6.6 \pm 2.2$ \\
        \multicolumn{1}{|l|}{\textbf{\LevelTwo}  (medium)}   & $6.6 \pm 2.3$   & $6.1 \pm 1.7$    & $6.8 \pm 2.4$   & $6.5 \pm 2.2$ \\
        \multicolumn{1}{|l|}{\textbf{\LevelThree} (high)}  & $6.7 \pm 2.5$   & $6.7 \pm 2.7$    & $7.4 \pm 3.0$   & $7.1 \pm 2.7$ \\
        \hline
    \end{tabular}
    \end{subtable}
    \hspace{4em}
    \begin{subtable}{.35\textwidth}
    \centering
    \caption*{(c') Complementary analysis}
    \label{tab:vertical_gaze_dispersion-complementary}
    \begin{tabular}{|c|c|}
        \hline
        \multicolumn{2}{|c|}{\textbf{Distraction}}\\
        \hline
        \textbf{Calculations}      & \textbf{No Calculations} \\    
        \hline
        $6.3 \pm 1.9$       & $7.1 \pm 2.5$ \\
        $6.2 \pm 2.3$       & $6.8 \pm 2.3$ \\
        $7.0 \pm 2.7$       & $7.5 \pm 2.8$ \\
        \hline
    \end{tabular}
    \end{subtable} 
\end{table*}

\subsubsection{Horizontal and Vertical Gaze Dispersions} 
For horizontal gaze dispersion, there was only a significant main effect of ACC use ($F(1,301.08)=65.531, p<.001$), with greater horizontal gaze dispersion when ACC was deactivated ($8.7 \pm 3.1$ globally) compared to when it was activated ($7.3 \pm 2.3$ globally) (see \Cref{tab:horizontal_gaze_dispersion-main}~(b)). Participants thus had a more concentrated horizontal gaze when ACC was engaged. There were no significant effects of cognitive distraction ($F(1,301.05)=.023, p=.880$) or driving environment complexity ($F(2,301.06)=.226, p=.798$), and no significant interactions.

For vertical gaze dispersion, significant main effects were found for cognitive distraction ($F(1,301.04)=9.645, p=.002$), driving environment complexity ($F(2,301.05)=3.974, p=.020$), and ACC use ($F(1,301.07)=4.232, p=.041$), with no significant interactions. 
\Cref{tab:vertical_gaze_dispersion-main}~(c) shows that vertical gaze dispersion was higher under cognitive distraction (distraction: $6.9 \pm 2.4$; no distraction: $6.5 \pm 2.3$, globally), and increased with driving environment complexity (\LevelOne: $6.5 \pm 2.2$; \LevelTwo: $6.5 \pm 2.1$; \LevelThree: $7.2 \pm 2.8$, globally). Finally, as with horizontal gaze dispersion, vertical gaze dispersion was greater when ACC was deactivated ($6.8 \pm 2.5$ globally), compared to when it was activated ($6.5 \pm 2.3$ globally). 

\subsection{Complementary analyses of the effects of mental calculation}
We conducted complementary analyses to further investigate the effects of cognitive distraction on gaze-based indicators. Indeed, we observed that participants directed their gaze more toward the road center area in driving sessions without a secondary cognitive task, whereas vertical gaze dispersion increased in sessions involving mental calculations. 
This contrasts with previous studies, which typically report a concentration of gaze on the road center area under cognitive distraction~\cite{Victor2005Sensitivity,Harbluk2007AnOnroad,Wang2014TheSensitivity}.
We hypothesized that this discrepancy may stem from differences in the nature of the cognitive tasks. Unlike the continuous n-back tasks commonly used in earlier studies, our task involved intermittent mental calculations. Notably, addition problems are associated with upward and rightward gaze shifts~\cite{Hartmann2015Spatial,Salvaggio2022ThePredictive}, which may account for the observed increase in vertical dispersion.

We thus segmented gaze data from the three sessions involving cognitive distraction, in each level of driving environment complexity, into (1)~segments during which participants were actively solving arithmetic problems (from the onset of the voice agent’s question to the participant’s response, lasting approximately 7 seconds: 2 seconds for the question and 3.5 ± 1.5 seconds for the response), and (2)~interleaving segments between calculations.  
We then computed the same gaze-based indicators as in the main analyses. This time, we were able to perform an ANOVA with repeated measures and compute effect sizes using partial eta squared~\cite{Richardson2011Eta}.

For percent road center, the analysis revealed no significant main effects of neither calculation phase ($F(1,28)=2.155, p=.153, \eta_p^2=.071$) nor driving environment complexity ($F(2,56)=1.553, p=.221, \eta_p^2=.053$), and no significant interaction. In contrast, for both horizontal and vertical gaze dispersions, calculation phase ($F(1,28)=24.808, p<.001, \eta_p^2=.470$ and $F(1,28)=12.364, p=.002, \eta_p^2=.306$, respectively) had significant effects with large effect sizes. No significant effects were found for driving environment complexity (horizontal: $F(2,56)=1.816, p=.172, \eta_p^2=.061$; vertical: $F(2,56)=3.016, p=.057, \eta_p^2=.097$), and there were no significant interactions.

\Cref{tab:PRC-complementary}~(a'), (b'), and (c') show that our initial hypothesis was incorrect. In fact, both horizontal and vertical gaze dispersions were lower during the calculation phase, indicating a more concentrated gaze toward the road center. The increased gaze dispersion observed in the main analyses for sessions with cognitive distraction stemmed from the interleaving periods between calculations, not from the calculations themselves. 
Interestingly, gaze dispersion (both horizontal and vertical) tended to be higher during non-calculation periods in sessions with distraction than in sessions without distraction. 
Conversely, horizontal gaze dispersion was lower during the calculation phase in sessions with distraction than in sessions without distraction. These results suggest that mental calculations prompted gaze concentration on the road center, while the periods between calculations were marked by more dispersed gaze patterns. 

\section{Discussion}
We analyzed two gaze-based indicators of driver cognitive distraction (percent road center and gaze dispersion) during a simulator study in three levels of driving environment complexity and two levels of cognitive distraction, where participants could freely activate or deactivate ACC. 

\textbf{Effects of cognitive distraction.} The results of the main analyses indicate that cognitive distraction significantly affects both percent road center and vertical gaze dispersion, but not horizontal gaze dispersion. The percent road center is lower during drives involving cognitive distraction, and the vertical gaze dispersion is higher. 
These findings contradict previous results reported in the literature. For instance, \citet{Victor2005Sensitivity} reported that cognitively distracted drivers increase their road viewing time and spatially concentrate their gaze in the road center area at the expense of peripheral glances.
One possible explanation for this discrepancy lies in the type of distracting task being used. Prior studies often used continuous tasks such as the n-back, which involves working memory. In contrast, our study used an intermittent mental calculation task, which is known to trigger upward gaze shifts during addition problems~\cite{Hartmann2015Spatial,Salvaggio2022ThePredictive}.

To further investigate the origin of our findings, we conducted complementary analyses focusing on the sessions with cognitive distraction. We compared the gaze-based indicators from driving segments occurring during mental arithmetic to those from the interleaving segments between calculations. 
It turned out that our initial hypothesis that the increase in vertical gaze dispersion was due to cognitive distraction was incorrect. In fact, participants showed more concentrated gaze both horizontally and vertically during the mental calculation segments. The increased gaze dispersion observed in the main analysis originated from the interleaving segments between calculations in the sessions with distraction, not the calculations themselves. Participants actually had more concentrated gaze toward the road center during calculations than in sessions without distraction, whereas between calculations, their gaze was more dispersed than in distraction-free sessions. One possible explanation is that, after a cognitively demanding episode leading to gaze concentration, drivers may engage in increased gaze dispersion to regain situation awareness of their driving surroundings. This hypothesis should be investigated in future studies.

\textbf{Effects of traffic.} The main analysis reveals that driving environment complexity only influences vertical gaze dispersion, but not the percent road center or horizontal gaze dispersion. The vertical gaze dispersion is higher in the most complex environment than in the least complex one. 
Previous studies found that, as driving task complexity increases, drivers increase their road viewing time and spatially concentrate their gaze on the road center area~\cite{Victor2005Sensitivity}.

\textbf{Effects of ACC.} Finally, we show that ACC use significantly influences all gaze-based indicators. When ACC is engaged, drivers look more often at the road center, and both horizontal and vertical gaze dispersions are lower compared to when ACC is disengaged. 
While \citet{Louw2017Are} reported more dispersed gaze (particularly horizontally) during SAE Level 2 automation than in manual driving, we may explain the discrepancies with our findings by differences in automation level. In our study, even with ACC engaged, drivers remain responsible for lateral control, which likely kept them more visually focused on the driving task.  
Importantly, participants were free to activate or deactivate ACC at any time, meaning that comparisons between ACC-on and ACC-off conditions do not necessarily involve the same road segments. This limits the ability to isolate the effect of ACC use from potential differences in driving context.

\section{Conclusion}
This study examined how driver cognitive distraction, driving environment complexity (specifically, traffic conditions), and adaptive cruise control (ACC) use influence gaze parameters.
A total of $N = 29$ participants completed six driving simulator scenarios, combining three levels of driving environment complexity and two levels of cognitive distraction, with the freedom to activate or deactivate ACC at any time. 

Our results show that vertical gaze dispersion increases in more complex driving environments, while ACC use leads to gaze concentration toward the road center with lower horizontal and vertical gaze dispersions. 
In cognitively distracting scenarios, drivers looked less at the road center and exhibited greater vertical gaze dispersion. 
However, complementary analyses revealed that the gaze was actually more concentrated toward the road center during mental calculations, and that the increased dispersion stemmed from the interleaving periods between calculations.

Future studies could more finely analyze how gaze dispersion varies during and after cognitive demand across diverse distracting tasks performed while driving. It would be worth examining the underlying causes and mechanisms at play.

\begin{acks}
The work by A. Halin and A. Deliège was supported by the SPW EER, Wallonia, Belgium under grant n°2010235 (ARIAC by \href{https://www.digitalwallonia.be/en/}{DIGITALWALLONIA4.AI}). 
The authors thank AISIN Europe for providing access to their driving simulator, and specifically acknowledge Richard Virlouvet and Frédéric Burguet for their assistance and support with the simulator.
\end{acks}

\bibliographystyle{ACM-Reference-Format}
\bibliography{main}

\end{document}